\documentclass{ws-ijmpd}

\begin{document}

\markboth{Domainko et al.}
{Galaxy clusters with H.E.S.S.}

%
\catchline{}{}{}{}{}
%

\title{NEW RESULTS FROM H.E.S.S. OBSERVATIONS OF GALAXY CLUSTERS}

\author{WILFRIED DOMAINKO}

\address{Max-Planck-Institute for Nuclear Physics, Saupfercheckweg 1\\
69117 Heidelberg
Germany\\
wilfried.domainko@mpi-hd.mpg.de}

\author{DALIBOR NEDBAL}

\address{Institute of Particle and Nuclear Physics, Charles University, V Holesovickach 2,\\
180 00 Prague 8, Czech Republic\\
nedbal@ipnp.troja.mff.cuni.cz}

\author{JAMES A. HINTON}

\address{School of Physics \& Astronomy, University of Leeds,\\
Leeds LS2 9JT, UK\\
j.a.hinton@leeds.ac.uk}

\author{OLIVIER MARTINEAU-HUYNH}

\address{LPNHE, Universit\'e Pierre et Marie Curie Paris 6, Universit\'e Denis Diderot
Paris 7, CNRS/IN2P3, 4 Place Jussieu, F-75252,\\ 
Paris Cedex 5, France\\
olivier.martineau-huynh@lpnhep.in2p3.fr}

\author{FOR THE H.E.S.S. COLLABORATION}

\maketitle

\begin{history}
\received{Day Month Year}
\revised{Day Month Year}
\comby{Managing Editor}
\end{history}

\begin{abstract}
Clusters of galaxies are believed to contain a significant population of cosmic rays. From the radio and probably hard X-ray bands it is known that clusters are the spatially most extended emitters of non-thermal radiation in the Universe. Due to their content of cosmic rays, galaxy clusters are also potential sources of VHE ($>$100 GeV) gamma rays. Recently, the massive, nearby cluster Abell 85 has been observed with the H.E.S.S. experiment in VHE gamma rays with a very deep exposure as part of an ongoing campaign. No significant gamma-ray signal has been found at the position of the cluster. The non-detection of this object with H.E.S.S. constrains the total energy of cosmic rays in this system. For a hard spectral index of the cosmic rays of -2.1 and if the cosmic-ray energy density follows the large scale gas density profile, the limit on the fraction of energy in these non-thermal particles with respect to the total thermal energy of the intra-cluster medium is 8\% for this particular cluster. This value is at the lower bounds of model predictions.
\end{abstract}


\section{Introduction}	

Galaxy clusters are the spatially most extended emitters of non-thermal radiation in the Universe. Radio observations show most significantly the presence of accelerated electrons in these systems\cite{giovannini00}\cdash\cite{gitti07}. Additionally, it is also expected that clusters contain a significant population of hadronic cosmic rays. Due to their spatial extension and due to the presence of magnetic fields in the $\mu$G range, clusters confine and accumulate cosmic ray protons with energies of up to $\sim10^{15}$ eV which were accelerated in the cluster volume\cite{voelk96}\cdash\cite{berezinsky97}.

Cosmic rays can be accelerated at several sites in galaxy clusters. Large-scale shock waves connected to cosmological structure formation are promising sources of cosmic rays \cite{colafrancesco00}\cdash\cite{ryu03}. Supernova remnant shocks and galactic winds have also the ability to produce high-energy particles \cite{voelk96}. Additionally, active galactic nuclei (AGNs) can distribute non-thermal particles in the cluster volume \cite{ensslin97}\cdash\cite{hinton07}.

A component of high energy particles should result in gamma-ray emission in galaxy clusters\cite{blasi07}. Hadronic cosmic rays can produce gamma rays through inelastic collisions with thermal protons and nuclei as targets and subsequent $\pi^0$ decay\cite{dennison80}$^,$\cite{voelk96}. Alternatively, leptonic cosmic rays with sufficiently high energies can up-scatter cosmic microwave background (CMB) photons to the gamma-ray range in inverse Compton processes \cite{atoyan00}\cdash\cite{gabici04}.

Despite the arguments for potential gamma-ray emission given above, no galaxy cluster has firmly been established as a gamma-ray source\cite{reimer03}. In the very-high energy gamma-ray range (VHE, E $>$ 100 GeV) upper limits have been reported for several clusters by the {\it Whipple}\cite{perkins06} and {\it H.E.S.S.} collaboration\cite{domainko07}\cdash\cite{aharonian09}. In this paper the results of observations of the galaxy cluster Abell~85 with the H.E.S.S. experiment are presented.

\section{The H.E.S.S. experiment}

The observations were performed with the H.E.S.S. telescope array, consisting of four imaging atmospheric Cherenkov telescopes located at the Khomas highlands in Namibia\cite{hinton04}. It has a field of view of $\sim$5$^\circ$ and observes in the VHE gamma-ray regime. The whole system is well suited to study galaxy clusters since due to its large field of view, H.E.S.S. can detect extended sources and it is expected that clusters feature extended VHE gamma-ray emission.

\section{Target Abell~85 and results}

Abell~85 is a nearby (z = 0.055) massive and hot (T $\approx$ 7 keV) galaxy cluster with a complex morphology\cite{kempner02}\cdash\cite{durret05}. It hosts a colling core at its center. In cooling core clusters, the central gas density is large enough that the radiative cooling time due to thermal X-ray emission of the intra-cluster gas is shorter than the age of the galaxy cluster. Additionally, it shows two sub-clusters merging with the main cluster which is quite uncommon for a cooling core cluster. Presumably the merging sub-clusters have not reached the central region of the main cluster and have therefore not disrupted the existing cooling core\cite{kempner02}.

This object has been observed with H.E.S.S. for 32.5 hours live time of good quality in October and November 2006 and in August 2007. The mean zenith angle of the observations was 18$^{\circ}$ which resulted in an energy threshold of 460 GeV. For the analysis of the data various assumptions on the spatial extension of the potential gamma-ray emission have been adopted\cite{aharonian09}. None of the probed regions showed a significant gamma-ray excess and hence upper limits have been derived. For obtaining the upper limits the approach of Feldman \& Cousins\cite{feldman98} assuming a spectral index of the emission of -2.1 was used. The first region for which an upper limit has been calculated is the high gas density core region. For a radius of 0.1$^{\circ}$ (0.4 Mpc at the object) around the cluster center a flux upper limit of F~($>$460~GeV)~$<$~3.9~$\times$~10$^{-13}$~ph.~cm$^{-2}$~s$^{-1}$ has been found. As a next area, a radius of 0.49$^{\circ}$  (1.9 Mpc) which corresponds to the size of the detected thermal X-ray emission of the cluster, has been investigated. Here the upper limit in VHE gamma-ray flux is F ($>$460 GeV) $<$ 1.5 $\times$ 10$^{-12}$ ph. cm$^{-2}$ s$^{-1}$. Finally a very extended region with a radius of 0.91$^{\circ}$ (3.5 Mpc) has been explored to search for any emission connected to the accretion shock of the cluster. For this case the data set is reduced to 8.6 live hours due to the lack of suitable off-source data for the background estimation and there the flux upper limit was determined to F~($>$460~GeV)~$<$~9.9~$\times$~10$^{-12}$ ph.~cm$^{-2}$~s$^{-1}$.

\begin{figure}[pt]
\centerline{\psfig{file=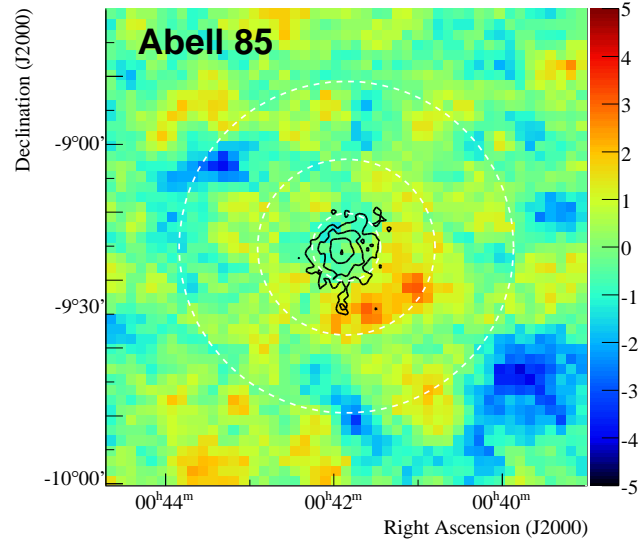,width=8.7cm}}
\vspace*{8pt}
\caption{Skymap of Abell~85 obtained with H.E.S.S. The black contours are from ROSAT PSPC X-ray observations. The dashed circles show radii of 0.4 Mpc, 1 Mpc and 1.9 Mpc. For details see main text.}
\end{figure}

\section{Discussion}

From the upper limits of the gamma-ray luminosity of the cluster Abell~85 it is possible to estimate upper limits on the total energy in hadronic cosmic rays in this cluster. For this purpose a spectral index of the cosmic rays of -2.1 is adopted and it is further assumed that the spatial distribution of the cosmic rays follows the large scale distribution of the gas density excluding the central cooling core. These conditions seem to be realistic since no losses of hadronic cosmic rays at relevant energies occur in clusters and therefore the hard source spectrum of cosmic rays should be seen\cite{voelk96}\cdash\cite{berezinsky97} and since magneto-hydrodynamic instabilities disfavor very centrally peaked distributions of cosmic rays\cite{parker66}\cdash\cite{chandran06}. With the adopted assumptions it is found that the total energy in cosmic rays is less than 8\% of the thermal energy of the intra-cluster medium\cite{aharonian09}. This value is at the lower bounds of model predictions. Similar results have very recently also been inferred from deep radio observations for the galaxy cluster Abell~521\cite{brunetti08}. Therefore it seems quite likely that the next generation of gamma-ray observatories like CTA will be necessary to detect galaxy clusters in the VHE gamma-ray regime.


\end{document}